\documentclass[aps,pre,reprint,groupedaddress,showpacs]{revtex4-1}

\usepackage[dvips]{graphicx}
\begin{document}

\title{Particle-Size Effects in the Formation of Bicontinuous Pickering Emulsions}

\author{M.~Reeves}
\author{A.~T.~Brown}
\author{A.~B.~Schofield}
\author{M.~E.~Cates}
\altaffiliation{DAMTP, Centre for Mathematical Sciences, University of Cambridge, CB3 0WA, United Kingdom}
\author{J.~H.~J.~Thijssen}
\email[]{j.h.j.thijssen@ed.ac.uk}
\homepage[]{http://www2.ph.ed.ac.uk/~jthijsse/}
\affiliation{SUPA School of Physics \& Astronomy, The University of Edinburgh, Edinburgh, EH9 3FD, United Kingdom}

\date{\today}

\begin{abstract}
We demonstrate that the formation of bicontinuous emulsions stabilized by interfacial particles (bijels) is more robust when nanoparticles rather than microparticles are used. Emulsification via spinodal demixing in the presence of nearly neutrally wetting particles is induced by rapid heating. Using confocal microscopy, we show that nanospheres allow successful bijel formation at heating rates two orders of magnitude slower than is possible with microspheres. In order to explain our results, we introduce the concept of mechanical leeway i.e.~nanoparticles benefit from a smaller driving force towards disruptive curvature. Finally, we suggest that leeway mechanisms may benefit any formulation in which challenges arise due to tight restrictions on a pivotal parameter, but where the restrictions can be relaxed by rationally changing the value of a more accessible parameter.
\end{abstract}


\maketitle

\section{Introduction}\label{sec:introduction}

The directed assembly of colloidal particles enables the design of novel soft materials with bespoke 3D architectures. The desired assembly route can be selected by adjusting the interparticle interactions. For example, the electrostatic interaction between oppositely charged particles can be tuned to obtain ionic colloidal crystals rather than irreversible aggregation \cite{Leunissen2005Nature}. An alternative approach employs templates to guide particle assembly towards a target structure. For instance, sedimentation of microparticles onto structured solid templates has been used to direct colloidal-crystal assembly \cite{Blaaderen1997Nature} and binary crystals of nanoparticles have been grown via liquid--air interfacial assembly \cite{Dong2010Nature}. In both cases, the interaction between the assembling particles and the template is crucial: pattern--lattice mismatches of $\sim10$\% already cause crystal defects and liquid subphase properties significantly affect crystal quality \cite{Blaaderen1997Nature,Dong2010Nature}.

\begin{figure}
\includegraphics{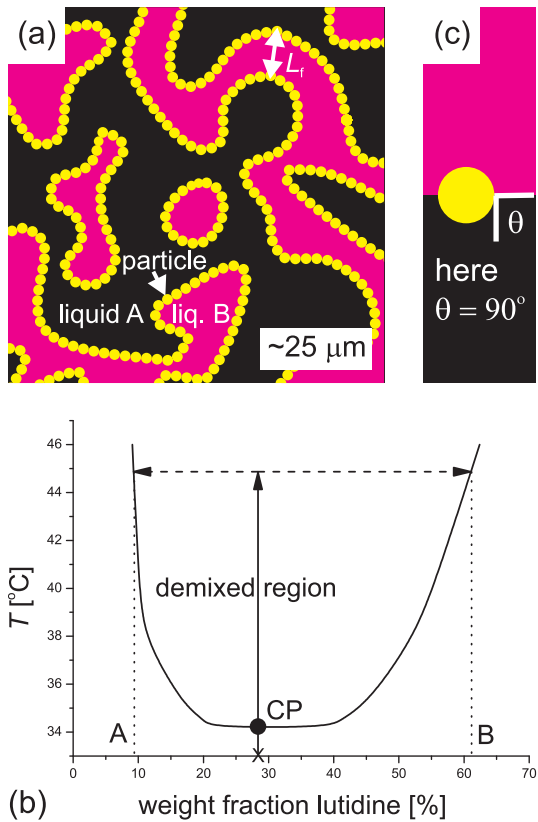}
\caption{(color online). (a) Schematic of a 2D slice through a 3D bijel: two continuous liquid channels (A black and B magenta), stabilized by a jammed layer of interfacial particles (yellow). $L_{\mathrm{f}}$: final channel width. (b) Coexistence curve for the water--lutidine (W--L) system (CP: critical point) \cite{Grattoni1993}. Vertical arrow: bijel formation, i.e.~a homogeneous mixture of W--L at the critical weight fraction (X) is heated from room temperature to $45 \ ^{\circ}\mathrm{C}$ (or $50 \ ^{\circ}\mathrm{C}$). Spinodal demixing results in two phases (A/B), with compositions given by the horizontal tie-lines. (c) Schematic of the contact angle $\theta$ (for $\theta = 90^{\circ}$).\label{fig:Figure_Introduction}}
\end{figure}

A startling case of liquid templating is the formation of bicontinuous Pickering emulsions \cite{Sanz2009PRL}, i.e.~bicontinuous interfacially jammed emulsion gels or bijels (Fig.~\ref{fig:Figure_Introduction}(a)) \cite{Stratford2005Science,Herzig2007NatMater,Firoozmand2009Langmuir,Cui2013Science,Tavacoli2015RSC}, which have been suggested for applications in fuel cells, microfluidics and tissue engineering \cite{Lee2010AdvMater,Wilson2006NatMater,Cates2008SoftMatter,Martina2005Biomaterials}. Bijel formation typically proceeds via spinodal demixing of a binary liquid containing colloidal particles (Fig.~\ref{fig:Figure_Introduction}(b)), which can arrest the phase separation by forming a jammed monolayer at the liquid--liquid interface. As in the cases discussed above, template--particle interactions are essential: bijels are only formed if the particles are (almost) neutrally wetting, otherwise emulsion droplets are formed \cite{Tavacoli2015RSC}. The parameter that quantifies this interaction is the contact angle $\theta$, which is a measure of the particle's position relative to the liquid interface: $\theta = 90^{\circ}$ is neutral wetting (Fig.~\ref{fig:Figure_Introduction}(c)). Unfortunately, tuning the mean value of $\theta$ is non-trivial and restraining its variance is harder still, making bijel formation challenging.

Ostensibly, reducing particle size $r$ given a fixed final bijel-channel width $L_{\mathrm{f}} \gg r$ (Fig.~\ref{fig:Figure_Introduction}(a)) would only make matters worse, as scaling down $r$ in a close-packed monolayer of particles with \emph{fixed} $\theta \ne 90^{\circ}$ requires a commensurate reduction in the local radius of curvature of the interface \cite{Kralchevsky2005Langmuir}. In other words, for a given non-neutrality, one might expect smaller particles to locally demand a more strongly curved interface and hence be more disruptive to bicontinuity on a chosen scale $L_{\mathrm{f}} \gg r$. However, this ignores the particle-size dependence of the stiffness of the particle-laden liquid interface, which might specifically aid small particles in overcoming off-neutral wetting.

In this paper, we experimentally explore the effect of particle size on bijel formation. We find that bijels are formed more robustly when nanoparticles rather than microparticles are used: nanospheres allow minimum heating rates two orders of magnitude slower than microspheres, with the latter stabilizing droplet emulsions rather than bijels at slow rates. We discuss our results in the context of mechanical leeway, i.e.~interfacial particles that are smaller lead to a less rigid interface between the two liquid phases, resulting in a smaller driving force towards disruptive curvature. Finally, we discuss the implications of leeway mechanisms in the (directed) self-assembly of functional formulations based on particle-template or even particle-particle interactions.

\section{Materials and Methods}\label{sec:materials_methods}

\subsection{Materials}\label{subsec:materials}

For particle synthesis, tetraethyl orthosilicate (TEOS, $\ge99$\%, Aldrich), 35\% ammonia solution (reagent grade, Fisher Scientific), ethanol absolute (VWR Chemicals), fluorescein isothiocynate (FITC, 90\% isomer 1, Aldrich) and (3-aminopropyl)triethoxysilane (APTES, 99\%, Aldrich) were used as received. For bijel preparation, 2,6-lutidine ($\ge99$\%, Aldrich) and Nile Red (Aldrich) were used as received; distilled water was run through a Milli-Q (Millipore) filtration system to perform deionization (to a resistivity of at least $12 \ \mathrm{M}\Omega\cdot\mathrm{cm}$).

Here, we formed (bicontinuous) Pickering emulsions by spinodal demixing of the binary liquid water-lutidine, heated at various rates in the presence of colloidal particles. Note that the water-lutidine (W-L) interfacial tension $\gamma_{\mathrm{WL}}$ is temperature-dependent and orders of magnitude lower than that of typical water-alkane systems. According to Ref.~\cite{Grattoni1993}, $\gamma_{\mathrm{WL}}$ ranges from $\sim0.01 \ \mathrm{mN}\:\mathrm{m}^{-1}$ at $34.2 \ ^{\circ}\mathrm{C}$ (just above the lower critical solution temperature of $34.1 \ ^{\circ}\mathrm{C}$) to $\sim0.4 \ \mathrm{mN}\:\mathrm{m}^{-1}$ at $46.0 \ ^{\circ}\mathrm{C}$. During slow heating at $1 \ ^{\circ}\mathrm{C}\:\mathrm{min}^{-1}$, it takes about 6 s to get from $34.1 \ ^{\circ}\mathrm{C}$ to $34.2 \ ^{\circ}\mathrm{C}$ and about 12 min to get to $46.0 \ ^{\circ}\mathrm{C}$.

\subsection{Particle Synthesis}\label{subsec:particle_synthesis}

\begin{figure}
\includegraphics{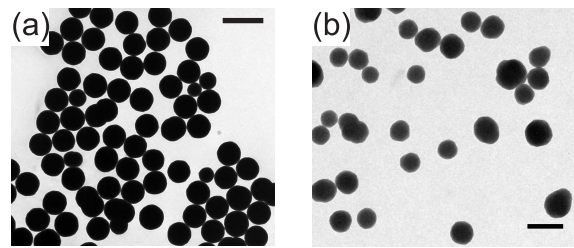}
\caption{Transmission Electron Microscopy (TEM) images of the St\"{o}ber silica (a) microparticles and (b) nanoparticles used in this study. The particles have an average radius of (a) 348 nm and (b) 63 nm; polydispersity (a) 6.4\% and (b) 15\% (analyzed images taken at different magnification to improve statistics). Scale bar: (a) $1 \ \mu\mathrm{m}$ and (b) 200 nm.\label{fig:Figure_Particle_Size}}
\end{figure}

The particles used in this study were synthesized using the St\"{o}ber method \cite{Stober1968JoCaIS}, modified to include the dye FITC via the linking molecule APTES \cite{Blaaderen1992Langmuir}. For the microparticles (MPs), a dye mixture of 0.584 g APTES, 0.107 g FITC and 4.0 mL ethanol was prepared overnight by stirring. The following day, a reaction mixture of 1.5 L ethanol, 186 mL 35\% ammonia solution and 60 mL TEOS was prepared, and the dye mixture added. The entire reaction mixture was kept in a refrigerator for 24 hours at $\sim10 \ ^{\circ}\mathrm{C}$. This resulted in particles with a radius of $0.36 \ \mu\mathrm{m}$ as measured by Dynamic Light Scattering (DLS) and $0.35 \ \mu\mathrm{m}$ according to Transmission Electron Microscopy (TEM, Fig.~\ref{fig:Figure_Particle_Size}(a)).

The nanoparticles (NPs) were synthesized in a similar fashion to the MPs, except the reaction temperature was $25 \ ^{\circ}\mathrm{C}$ and the concentration of dye mixture was increased to take account of the increase in surface-to-volume ratio which accompanies a decrease in particle radius. This is an important consideration, as it has been shown that the presence of APTES on the silica surface is crucial for meeting the neutral-wetting requirement in the W-L system \cite{White2008JPCM,White2011JoCaIS} -- it has been suggested that the surface decorations act to disrupt the wetting layer of lutidine which spontaneously forms around the particles when approaching the phase separation temperature \cite{Beysens1985PRL,Sluckin1990PRA}. For the NPs, DLS returned a particle radius of $0.08 \ \mu\mathrm{m}$ and TEM returned $0.063 \ \mu\mathrm{m}$ with a polydispersity of 15\% (Fig.~\ref{fig:Figure_Particle_Size}(b)). We have confirmed that the NPs ($1.51 \pm 0.06 \ \mathrm{g}\:\mathrm{mL}^{-1}$) have a lower density than the MP ($1.63 \pm 0.03 \ \mathrm{g}\:\mathrm{mL}^{-1}$) (density meter, Anton Paar, DMA 4500), presumably due to the higher dye concentration by volume \cite{Blaaderen1993JoCaIS}, which could lead to enhanced shrinkage in the vacuum of the TEM \cite{Imhof1999JPCB}. The $4.5\times$ decrease in DLS particle size closely matches the $4.5\times$ increase in APTES concentration compared to the MP synthesis, so the NPs and MPs are expected to have identical surface chemistries.

To remove excess APTES and FITC from the synthesis product, the particles were washed by repeated centrifugation/redispersion: $2\times$ ethanol then $4\times$ water for the MPs and $2\times$ ethanol then $8\times$ water for the NPs. Subsequently, the particles were pre-dried at room temperature in a fume hood and ground with a mortar and pestle. Prior to sample preparation, particles were dried at 20 mBar and $170 \ ^{\circ}\mathrm{C}$ (no more than $100 \ \mathrm{mg}$ per vial and no more than 3 vials at the same time) \cite{White2011JoCaIS}. This removes surface-bound water and may cause moderate dehydroxylation of the silica surface \cite{Zhuravlev2000CaSA}. The drying time was tuned to optimize bijel quality as assessed by visual inspection of confocal micrographs; dried particles were stored in a desiccator in the presence of a silica gel.

\subsection{Sample Preparation}\label{subsec:sample_preparation}

First, dried particles were dispersed in deionized water by ultrasonication (Sonics VibraCell). The MPs were sonicated for $\left( 2\times2 \right)$ minutes at 8 W with $\left( 2\times10 \right)$ s of vortex mixing in between. To ensure proper redispersion, NPs were additionally sonicated for $\left( 1\times10 \right)$ minutes at 8 W and vortex mixed for $\left( 1\times10 \right)$ s. Lutidine was then added to give a mixture with a critical composition, i.e.~a mass ratio of W:L = 72:28 (Fig.~\ref{fig:Figure_Introduction}(b)) \cite{Grattoni1993}, so that spinodal decomposition would be (at least initially) the preferred phase separation mechanism. To allow confocal imaging of the lutidine-rich phase, the fluorescent dye Nile Red had been added to the lutidine at a concentration of around $10 \ \mu\mathrm{M}$ (we checked that Nile Red partitions into the lutidine-rich phase and that concentrations as low as $1 \ \mu\mathrm{M}$ gave similar bijels). The sample mixture was transferred to a glass cuvette (Starna 21-G-1 with pathlength 1 mm) and placed inside a metal block, which was itself placed inside a temperature stage (Instec, TSA02i). Emulsification via liquid-liquid demixing was initiated by heating the sample to a target temperature above the lower critical solution temperature (LCST) of $34 \ ^{\circ}\mathrm{C}$.

Slow heating ($\le5 \ ^{\circ}\mathrm{C}\:\mathrm{min}^{-1}$) was achieved by programming the temperature stage to ramp the temperature $T$ at the desired rate $\dot{T}$, from room temperature ($\approx20 \ ^{\circ}\mathrm{C}$) to $45 \ ^{\circ}\mathrm{C}$. Heating rates were extracted from the $\left( T,t \right)$-graphs produced by the stage software and we have used a thermocouple to ascertain that at these slow rates the sample temperature does not lag the stage temperature; estimate of corresponding error in heating rate $\sigma_{\dot{T}} = 0.1 \ ^{\circ}\mathrm{C}\:\mathrm{min}^{-1}$. For a heating rate of $17 \ ^{\circ}\mathrm{C}\:\mathrm{min}^{-1}$, we adopted a method from Ref.~\cite{Herzig2007NatMater}: the temperature stage and metal block were pre-warmed to $45 \ ^{\circ}\mathrm{C}$ or $50 \ ^{\circ}\mathrm{C}$ and the room-temperature cuvette was inserted. We have confirmed this heating rate by measuring the time it took to reach phase separation at the LCST of $34 \ ^{\circ}\mathrm{C}$ from room temperature; estimate of corresponding error in heating rate $\sigma_{\dot{T}} = 3 \ ^{\circ}\mathrm{C}\:\mathrm{min}^{-1}$. For higher heating rates, the cuvette was placed on top of a small cardboard box (to prevent thermal conduction away from the cuvette) inside a microwave (DeLonghi, P80D20EL-T5A/H, 800 W, set to ``auto-defrost 100 g'' i.e.~40\%) \cite{Lee2010AdvMater}. The sample was irradiated for 5 s (or 6 s) and then quickly transferred to the temperature stage at $50 \ ^{\circ}\mathrm{C}$. We have checked by visual inspection that the sample remained opaque (i.e.~phase separated) upon transfer from the microwave to the temperature stage. The corresponding heating rate was calculated as $\left( 50 \ ^{\circ}\mathrm{C} - 20 \ ^{\circ}\mathrm{C} \right) / 5 \ \mathrm{s} = 360 \ ^{\circ}\mathrm{C}\:\mathrm{min}^{-1}$, with an estimated error of $30 \ ^{\circ}\mathrm{C}\:\mathrm{min}^{-1}$.

\subsection{Characterization \& Image Analysis}\label{subsec:characterization_image_analysis}

During or after emulsification, samples were imaged using fluorescence confocal microscopy. Fluorescence excitation was provided by a 488 nm laser (for FITC) and a 555 nm laser (for Nile Red); emission filters were used as appropriate. The two liquid domains could be distinguished by detecting the fluorescence of the Nile Red, while the location of the particles could be determined by detecting the fluorescence of the FITC.

To extract the bijel channel width $L$ from 2D confocal microscopy images, a pixel-based correlation function algorithm was run on the Nile-Red channel using Matlab. The algorithm constructs a radial distribution function $g\left( r \right)$ by multiplying pairs of pixel intensities, plotting the values against the distance between the pixels, and then taking an average; the bijel channel width or characteristic length scale is then taken to be the location of the first minimum in the plotted $g\left( r \right)$ \cite{Chaikin1995}. For the final bijel-channel width $L_{\mathrm{f}}$, this process was repeated on at least three separate images of the same bijel sample and an average was taken. The standard deviation of measurements made on several images of the same sample was taken as the error $\sigma_{L} \approx 3 \ \mu\mathrm{m}$.

\section{Results}\label{sec:results}

\begin{figure}
\includegraphics{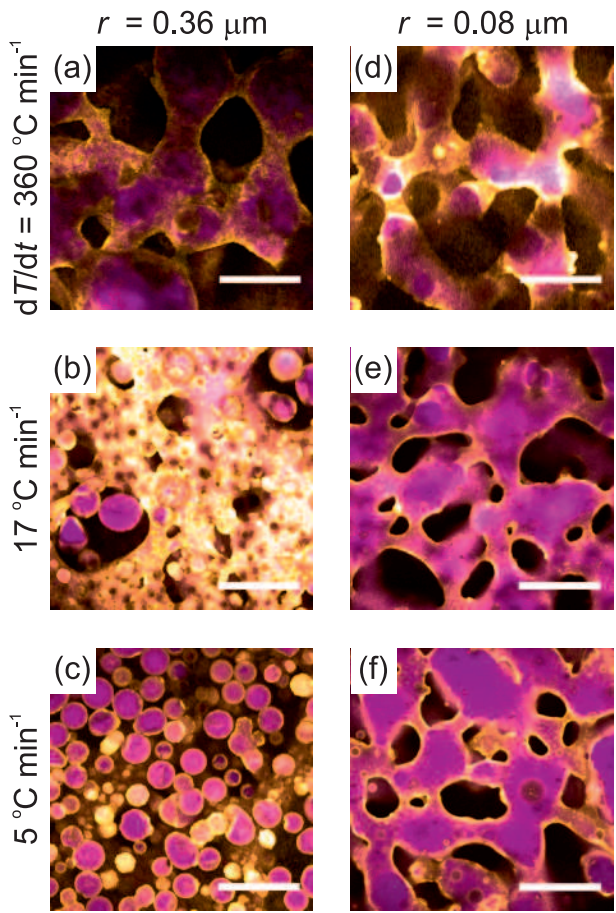}
\caption{(color online). Fluorescence confocal micrographs of final-state emulsions of water and lutidine (magenta), formed using various heating rates ($\mathrm{d}T / \mathrm{d}t$), stabilized by (nearly) neutrally wetting particles (yellow) of radius $r$. Particle volume fraction is (a) 2.6\%, (b--c) 2.2\% and (d--f) 0.7\%. Estimated relative error in heating rate $\sigma_{\dot{T}}<18$\% (Sec.~\ref{subsec:sample_preparation}). Scale bars: $100 \ \mu \mathrm{m}$. See Appendix \ref{sec:sample_homogeneity} for sample homogeneity.\label{fig:Figure_Confocal_Final}}
\end{figure}

We begin by comparing (bicontinuous) Pickering emulsions formed by spinodal decomposition of W-L mixtures, containing either nanoparticles (NPs) or microparticles (MPs), upon heating at various rates (Fig.~\ref{fig:Figure_Introduction}(b)). Fig.~\ref{fig:Figure_Confocal_Final} presents a confocal-microscopy overview of the structures obtained for two different particle radii and three different heating rates. In all panels, the fluorescently labeled particles (yellow) appear at the liquid-liquid interface between the water-rich phase (black) and the fluorescently labeled lutidine-rich phase (magenta). Samples prepared with MPs show bicontinuous structures only for fast heating (Fig.~\ref{fig:Figure_Confocal_Final}(a)), whereas slow heating results in discrete droplets (Fig.~\ref{fig:Figure_Confocal_Final}(c)). In contrast, NPs invariably yield a percolating interface with both signs of curvature (Fig.~\ref{fig:Figure_Confocal_Final}(d-f)), which is an imperative characteristic of a bijel; note that slow heating with NPs (Fig.~\ref{fig:Figure_Confocal_Final}(e,f)) seems to yield a relatively higher number of thin necks compared to fast heating (Fig.~\ref{fig:Figure_Confocal_Final}(d)).

\begin{figure}
\includegraphics{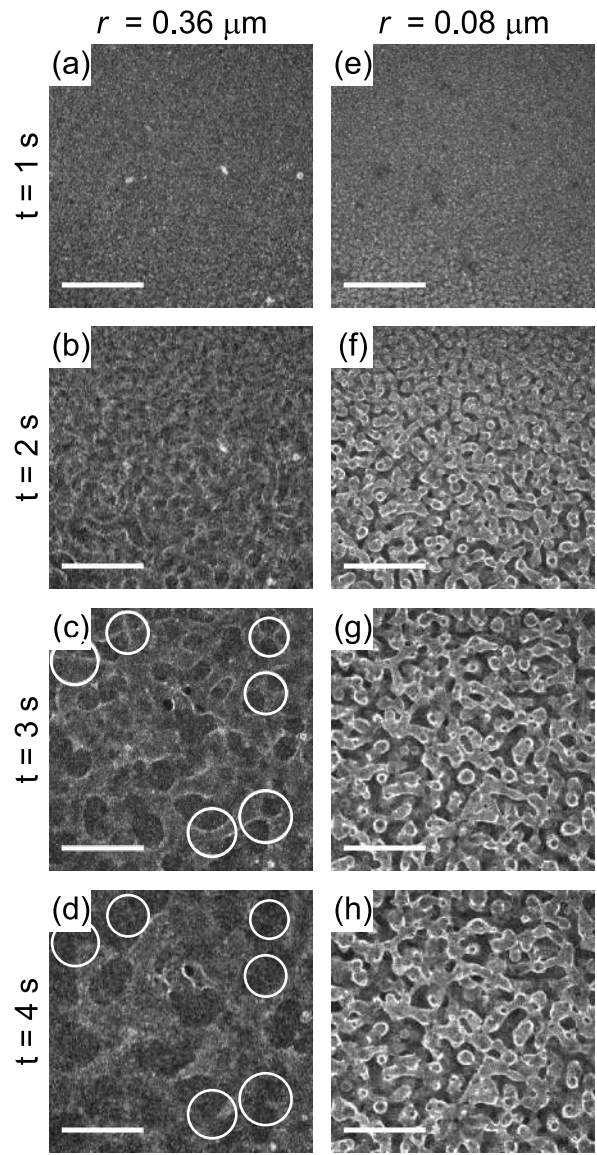}
\caption{Time sequences of confocal fluorescence micrographs showing water--lutidine mixtures containing (nearly) neutrally wetting particles of radius $r$ (white) during slow heating ($1 \ ^{\circ} \mathrm{C} \: \mathrm{min}^{-1}$). Particle volume fraction $\varphi$ is (a--d) 2.1\% and (e--h) 1.8\%. Note (c,d) the depercolation via (encircled) pinch-off events and (e--h) the formation of a bijel (also verified down to $\varphi_{\mathrm{NP}} = 0.7$\% (Appendix \ref{sec:np_low_volume_fraction})). Scale bars: $100 \ \mu \mathrm{m}$.\label{fig:Figure_Confocal_Formation}}
\end{figure}

Next, we compare the kinetics of bijel formation using MPs vs NPs, to explain the discrepancy in the structures obtained after slow heating (Fig.~\ref{fig:Figure_Confocal_Final}). Fig.~\ref{fig:Figure_Confocal_Formation} shows selected confocal micrographs from time-series recorded during slow heating in the presence of MPs vs NPs. Using MPs (Fig.~\ref{fig:Figure_Confocal_Formation}(a-d)), the interconnected domains present at $t = 2 \ \mathrm{s}$ have pinched off by $t = 4 \ \mathrm{s}$, resulting eventually in particle-stabilized droplets. By contrast, when using NPs (Fig.~\ref{fig:Figure_Confocal_Formation}(e-h)), connectivity is maintained until the structure is arrested, resulting in a bijel. Though we observe thinning of necks, we cannot find a convincing pinch-off event between Fig.~\ref{fig:Figure_Confocal_Formation}(g) and Fig.~\ref{fig:Figure_Confocal_Formation}(h). Note that we have also observed droplet formation via secondary phase separation (Appendix \ref{sec:secondary_nucleation}), but this does not seem to be a pivotal effect, i.e.~it can both happen and fail to happen irrespective of bijel formation failing or succeeding \cite{Herzig2007NatMater,White2011JoCaIS,Tavacoli2011AdvFunctMater,Witt2013SoftMatter}. This suggests that MPs fail to produce bijels via slow heating, because depercolation via pinch-off events occurs before the interfacial particles jam and lock-in the bicontinuous structure.

\begin{figure}
\includegraphics{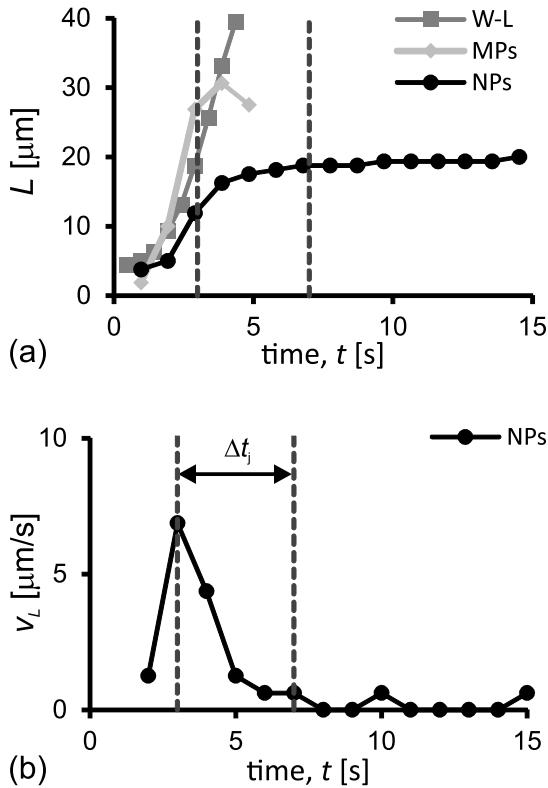}
\caption{(a) Measured channel width $L$ vs time $t$ during spinodal demixing upon heating at $1 \ ^{\circ}\mathrm{C}\:\mathrm{min}^{-1}$ of a critical mixture of water--lutidine without (W--L) and with (MPs) $0.36 \ \mu\mathrm{m}$ radius microparticles or (NPs) $0.08 \ \mu\mathrm{m}$ radius nanoparticles, the latter resulting in a bijel. (b) Corresponding coarsening speed $v_{L}$ for the NP data. The dashed vertical lines enclose the jamming time $\Delta t_{\mathrm{j}} \approx 4 \ \mathrm{s}$. Estimated error in $L$ is $\sigma_{L} \approx 3 \ \mu\mathrm{m}$.\label{fig:Figure_Coarsening}}
\end{figure}

To quantify the coarsening observed in Fig.~{\ref{fig:Figure_Confocal_Formation}}, we used image analysis to extract the channel width $L$ (Sec.~\ref{subsec:characterization_image_analysis}). Fig.~\ref{fig:Figure_Coarsening}(a) shows that the coarsening in the presence of MPs is similar to coarsening without particles, until $t = 5 \ \mathrm{s}$ when the bicontinuous structure has failed and MP-stabilized droplets have appeared. Coarsening in the presence of NPs initially follows the behavior of the W-L mixture without particles, but then levels off. As bijel formation at $1 \ ^{\circ}\mathrm{C}\:\mathrm{min}^{-1}$ fails with MPs (and without particles), Fig.~\ref{fig:Figure_Coarsening}(b) only shows the coarsening speed
\begin{equation}\label{eq:coarsening_speed}
v_{L} = \frac{L_{t_{\mathrm{i}}} - L_{t_{\mathrm{i-1}}}}{t_{\mathrm{i}} - t_{\mathrm{i-1}}}
\end{equation}
in the case of NPs; note that $v_{L}$ goes through a maximum at $t \approx 3 \ \mathrm{s}$ and is (more or less) 0 after $t \approx 7 \ \mathrm{s}$. As discussed below, we refer to the time between the maximum in $v_{L}$ and its levelling off as the `jamming time' $\Delta t_{\mathrm{j}}$.

\section{Discussion}\label{sec:discussion}

Having presented our experimental results, we first discuss how bijel formation can fail and how particles with off-neutral wetting can promote bijel failure. Simulations of spinodal demixing without particles in 3D, in the viscous hydrodynamic (VH) regime relevant here, have shown that depercolation proceeds via thinning of liquid channels followed by pinch-off events \cite{Pagonabarraga2001NewJournalPhysics}. Neutrally wetting particles can halt the demixing by attaching to and jamming at the liquid interface \cite{Stratford2005Science,Herzig2007NatMater,Tavacoli2015RSC}. However, off-neutral particles induce a spontaneous curvature $C_0$ when attached to liquid interfaces \cite{Kralchevsky2005Langmuir,Planchette2012SoftMatter,Jansen2011PRE,White2011JoCaIS}. This is because they are pushed together as coarsening decreases the interfacial area, while the interparticle contacts are not situated at the liquid interface (where they would be for $\theta = 90^{\circ}$). As bijels have empirically been shown to feature average mean curvature $\left< H \right> = 0$ \cite{Lee2010AdvMater,Reeves2015InPrep}, any $C_0 \ne 0$ is expected to disrupt bijel formation.

Note that secondary nucleation, i.e.~the formation of new droplets during spinodal decomposition, was not observed in the above-mentioned simulations \cite{Pagonabarraga2001NewJournalPhysics,Stratford2005Science,Jansen2011PRE}, presumably because the quench was instantaneous \cite{Witt2013SoftMatter}. Secondary nucleation during bijel formation has previously been observed in experiments and attributed to the finite rate of temperature change \cite{Herzig2007NatMater,Witt2013SoftMatter}. However, it has not been suggested that secondary nucleation is responsible for bijel failure, rather it results in droplets inside bijel channels \cite{Herzig2007NatMater} or even droplet-reinforced channels \cite{Witt2013SoftMatter}.

Intriguingly, our results show that bijel formation fails during slow heating with MPs, whereas it succeeds with NPs that were designed to have similar wetting. The NP contact angle could simply be closer to $90^{\circ}$. However, this does not agree with our observation that NPs allow bijel formation over a wider range of drying times, which is expected to correspond to a wider range of contact angles \cite{White2011JoCaIS}. Our fluorescence confocal time series suggest that MP bijels fail due to depercolation via pinch-off events. Pinch-off events may also occur for NPs: they can even be observed in 3D simulations of successful bijel formation \cite{Stratford2005Science}. However, we suggest that NPs sufficiently suppress the number of pinch-off events to allow successful bijel formation.

In order to explain why NPs facilitate bijel formation, we have found it particularly illuminating to consider the particle-size dependence of the ``driving force'' $F$ towards $C_0$ (Appendix \ref{sec:timescales}) i.e.~away from $\left< H \right> = 0$ for bijels \cite{Lee2010AdvMater,Reeves2015InPrep}. The bending-energy density of the particle-laden interface is
\begin{equation}\label{eq:bending_energy_density}
w = 2 \kappa \left( H - C_0 \right)^2 \ ,
\end{equation}
where $\kappa$ is the effective bending modulus of the interface \cite{Canham1970,Helfrich1973}, so
\begin{equation}\label{eq:bending_force}
F = \frac{\partial w}{\partial H} = -4 \kappa \left( C_0 - H \right) \ .
\end{equation}
Dimensional analysis suggests that $C_0 \propto -1/r$ and $\kappa \propto \gamma_{\mathrm{WL}} r^2$, which is backed by analytical calculations for spheres on a spherical cap \cite{Kralchevsky2005Langmuir}. As here $L_{\mathrm{f}} \gg r$, and so $\left| H \right| \sim 1/L_{\mathrm{f}} \ll \left| C_0 \right|$, we approximate Eq.~(\ref{eq:bending_force}) as
\begin{equation}\label{eq:bending_force_scaling}
F \approx -4 \kappa C_0 \propto \gamma_{\mathrm{WL}} r \ .
\end{equation}
Thus, NPs demand a more strongly curved interface ($C_0 \propto -1/r$), but the driving force towards that curvature is smaller ($F \propto r$).

To assess to what degree a smaller driving force can facilitate bijel formation, we compare the disruption time $\Delta t_{\mathrm{d}}$ to the jamming time $\Delta t_{\mathrm{j}}$; bijel formation can succeed if $\Delta t_{\mathrm{j}} < \Delta t_{\mathrm{d}}$. For the NPs, we can estimate the jamming time from Fig.~\ref{fig:Figure_Coarsening}(b). We define the jamming time as $\Delta t_\mathrm{j}=t_\mathrm{f}-t_\mathrm{in}$, where $t_\mathrm{in}$ is the time at which the jamming starts causing a decrease in the coarsening speed $v_L$ - the peak in Fig.~\ref{fig:Figure_Coarsening}(b) - and $t_\mathrm{f}$ is the time just before $v_L$ drops to zero. This gives $\Delta t_\mathrm{j, NP}\approx 4 \mathrm{s}$ at a heating rate of $1 \ ^{\circ}\mathrm{C}\:\mathrm{min}^{-1}$.

We cannot obtain the MP jamming time directly, since MP bijels fail at $1 \ ^{\circ}\mathrm{C}\:\mathrm{min}^{-1}$. However, we expect the jamming dynamics to be dominated by the instantaneous area fraction of interfacial particles, which is independent of particle radius, as long as the final lengthscale is fixed (Appendix \ref{sec:timescales}). As $L_{\mathrm{f}}\left(\mathrm{MP}\right) > L_{\mathrm{f}}\left(\mathrm{NP}\right)$ in Fig.~\ref{fig:Figure_Confocal_Formation}, i.e.~$\varphi_{\mathrm{NP}}$ is 1.8\% vs 0.7\% in Fig.~\ref{fig:Figure_Confocal_Final} and $L_{\mathrm{f}} \propto r / \phi$ \cite{Cates2008SoftMatter}, we expect $\Delta t_{\mathrm{j,MP}}>\Delta t_\mathrm{j,NP}\approx 4 \ \mathrm{s}$.

Conversely, we can estimate the MP disruption time, $\Delta t_{\mathrm{d,MP}} \approx 2 \ \mathrm{s}$ from the time of occurrence of pinch-off in confocal images (Fig.~\ref{fig:Figure_Confocal_Formation}(b--d)), whereas we cannot estimate $\Delta t_{\mathrm{d,NP}}$ because bijel formation succeeds here for NPs. However, we can predict the scaling of $\Delta t_{\mathrm{d}}$ with particle radius by balancing the driving (Eq.~(\ref{eq:bending_force_scaling})) and viscous-drag forces, to give
\begin{equation}\label{eq:disruption_time}
\Delta t_{\mathrm{d}} \propto \frac{\eta \lambda^2}{\gamma_{\mathrm{WL}} r} \ ,
\end{equation}
where $\eta$ is a bulk fluid viscosity and $\lambda \gg r$ is the typical length scale of the disruption, which is independent of particle radius (Appendix \ref{sec:timescales}). Given that the MPs are $4.5\times$ larger than the NPs, and assuming effects of particle polydispersity and roughness are negligible, the inverse scaling of $\Delta t_{\mathrm{d}}$ with radius implies $\Delta t_{\mathrm{d},\mathrm{NP}} \approx 9 \ \mathrm{s}$.

These time-scale estimates help to explain the observed patterns of bijel failure, i.e.~they explain why ${\Delta t_{\mathrm{j,NP}} < \Delta t_{\mathrm{d,NP}}}$ but $\Delta t_{\mathrm{j,MP}} > \Delta t_{\mathrm{d,MP}}$. To account for any possible dependence of $\Delta t_{\mathrm{d}}$ on the (final) channel width, we have also verified that bijel formation is successful with NPs at $1 \ ^{\circ}\mathrm{C}\:\mathrm{min}^{-1}$ for similar $L_{\mathrm{f}}$, i.e.~for $\phi_{\mathrm{NP}} = 0.7$ vol-\% (Appendix \ref{sec:np_low_volume_fraction}). It is worth noting here that, based on the scaling proposed in Eq.~({\ref{eq:disruption_time}}), we had expected that bijel formation would succeed with MPs at $5 \ ^{\circ}\mathrm{C}\:\mathrm{min}^{-1}$. This is because it succeeds for similar $L_{\mathrm{f}}$ with NPs at $1 \ ^{\circ}\mathrm{C}\:\mathrm{min}^{-1}$ (and the NPs are about $5\times$ smaller than the MPs). As bijel formation with MPs is only barely successful at a higher rate of $17 \ ^{\circ}\mathrm{C}\:\mathrm{min}^{-1}$, this suggests that an additional mechanism might be at play here; currently planned simulations and experiments may be able to address this in the future. Having said that, the mechanical-leeway mechanism proposed here does point in the right direction, i.e.~it can explain why bijel formation is more robust when using NPs rather than MPs.

As shown above, slow heating increases the importance of bypassing droplet formation. We have suggested here that NPs succeed in this because of their larger mechanical leeway, whereas MPs may fail under similar conditions (Fig.~\ref{fig:Figure_Confocal_Final}). This also has technological relevance, since fast and homogeneous heating is challenging to achieve, putting severe restrictions on the choice of sample geometry and starting materials \cite{Tavacoli2015RSC}. Therefore, reducing particle size could greatly facilitate formulation, especially when tuning particle surface chemistry is non-trivial (as is often the case), even though a naive expectation based on an optimal (static) wetting geometry would suggest exactly the opposite trend.

This mechanical-leeway mechanism not only applies to bijels but to any liquid template for solid particles. More broadly, leeway mechanisms may well aid any formulation where challenges arise due to tight restrictions on a pivotal parameter, but where the restrictions can be relaxed by changing a more accessible parameter (here: particle size). This has important implications for the development of fabrication routes for advanced functional materials based on external templates. Moreover, it is potentially relevant to the design of any soft material with a bespoke architecture by adjusting particle interactions, e.g.~crystallization of spheres with a size variation above the hard-sphere crystallization threshold ($\sim 12\%$) \cite{Auer2001Nature} is possible by changing the ionic strength of the suspending medium so that the interparticle-interaction range is large enough \cite{Leunissen2007PNAS}.

\section{Conclusions}\label{sec:conclusions}

We have shown that the formation of bicontinuous Pickering emulsions (bijels) via liquid-liquid demixing is more robust with nanoparticles than with microparticles: a wider range of heating rates can be used. In addition, our results suggest that bijel formation using microparticles fails at low rates because the bicontinuous structure decays into discrete droplets via pinch-off events. To explain our observations, we have argued that interfacial microparticles with off-neutral wetting induce disruptive curvature, while nanoparticles of similar wetting benefit from a mechanical-leeway mechanism. In short, smaller particles give a smaller driving force towards disruptive curvature.

\begin{acknowledgments}
M.R.~is grateful to EPSRC for funding his PhD studentship. M.E.C.~is funded by the Royal Society. J.H.J.T.~acknowledges the Royal Society of Edinburgh/BP Trust Personal Research Fellowship for funding and The University of Edinburgh for awarding a Chancellor's Fellowship. The authors are also grateful for financial support from EPSRC EP/J007404. Thanks to Paul Clegg, Michiel Hermes and Alexander Morozov for useful discussions.
\end{acknowledgments}

\clearpage

\appendix

\section{Sample homogeneity}\label{sec:sample_homogeneity}

In this Appendix, we present several fluorescence confocal micrographs of a MP and a NP stabilized bijel, to demonstrate sample homogeneity.

\begin{figure}[h!]
\includegraphics{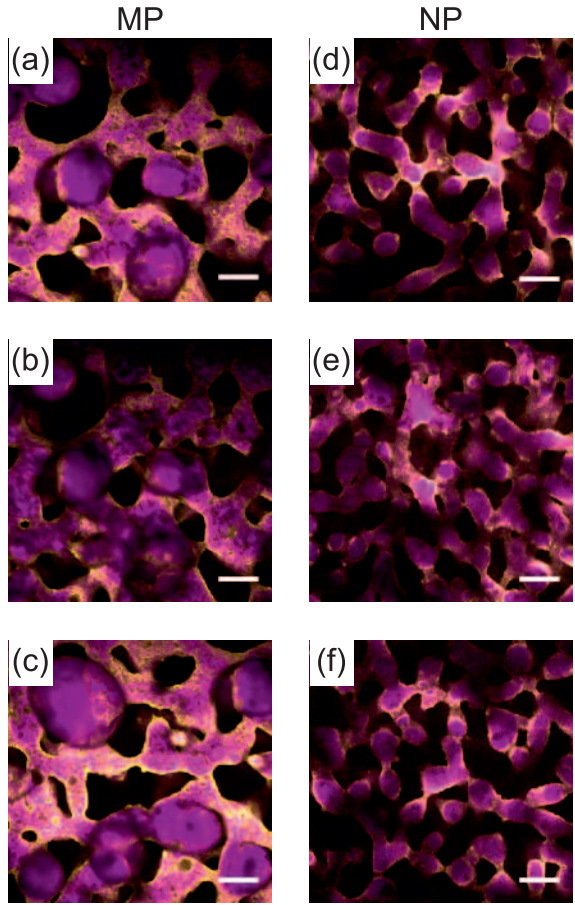}
\caption{(color online). Fluorescence confocal microscopy on bijels of water and lutidine (magenta), stabilized by nearly neutrally wetting particles (yellow), formed using microwave heating. Micrographs of a (a--c) microparticle (MP) and (d--f) nanoparticle (NP) stabilized bijel at three different positions (randomly chosen). Particle volume fraction is (a--c) 2.6\% and (d--f) 0.7\%. Scale bars: $100 \ \mu \mathrm{m}$.\label{fig:Figure_Sample_Homogeneity_Comb_sb100mu}}
\end{figure}

\newpage

\section{Secondary nucleation}\label{sec:secondary_nucleation}

Below are confocal micrographs, corresponding to Fig.~\ref{fig:Figure_Confocal_Formation}, to illustrate secondary nucleation.

\begin{figure}[h!]
\includegraphics{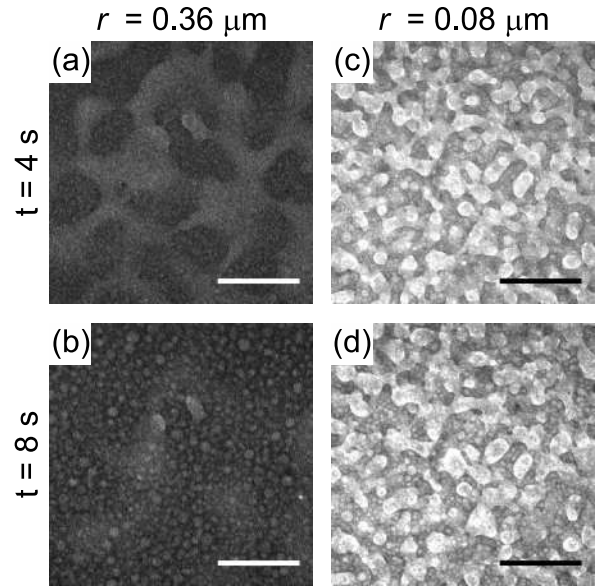}
\caption{Confocal fluorescence micrographs from two time-series showing water-lutidine(white) mixtures containing (nearly) neutrally wetting particles of radius $r$ during slow heating ($1 \ ^{\circ}\mathrm{C}\:\mathrm{min}^{-1}$). Particle volume fraction is (a,b) 2.1\% and (c,d) 1.8\%. Note that droplets have appeared, presumably due to secondary nucleation, which has previously been observed during slow quenches \cite{Herzig2007NatMater,Witt2013SoftMatter}. Scale bars: $100 \ \mu\mathrm{m}$.\label{fig:Figure_Secondary_Nucleation}}
\end{figure}

\newpage

\section{Slow heating using NPs at 0.7 vol-\%}\label{sec:np_low_volume_fraction}

Here, we present a confocal micrograph demonstrating successful bijel formation at a heating rate of $1 \ ^{\circ}\mathrm{C} \: \mathrm{min}^{-1}$ and a nanoparticle volume fraction of 0.7\% (see caption of Fig.~\ref{fig:Figure_Confocal_Formation} and discussion after Eq.~(\ref{eq:disruption_time})).

\begin{figure}[h!]
\includegraphics{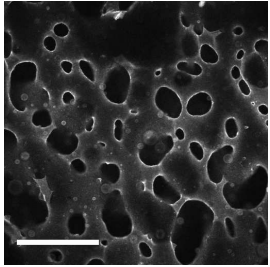}
\caption{Fluorescence confocal micrograph of a water-lutidine bijel, formed using a heating rate of $1 \ ^{\circ}\mathrm{C} \: \mathrm{min}^{-1}$, stabilized by nanoparticles of radius $0.08 \ \mu\mathrm{m}$ (white) at a volume fraction of 0.7\%. Scale bar: $100 \mu\mathrm{m}$.\label{fig:Figure_NP_Low_Phi}}
\end{figure}

\newpage

\section{Timescales}\label{sec:timescales}

In this Appendix, we obtain approximate scaling relationships for the timescales of jamming and disruption during bijel formation.

\subsection{Disruption Time}\label{subsec:disruption_time}

Following Canham and Helfrich \cite{Canham1970,Helfrich1973}, we start with the bending-energy density of a membrane
\begin{equation}\label{eq:bending_energy_density_full}
w = 2 \kappa \left( H - C_0 \right)^2 + \kappa_{\mathrm{G}} K \ ,
\end{equation}
in which $\kappa$ is the bending modulus, $H$ the mean curvature, $C_0$ the spontaneous curvature, $\kappa_{\mathrm{G}}$ the Gaussian bending modulus and $K$ the Gaussian curvature. Assuming the topology of the surface does not change substantially during the crucial stages of bijel formation, we omit the $K$ term \cite{Nakahara1990}:
\begin{equation}\label{eq:bending_energy_density_app}
w = 2 \kappa \left( H - C_0 \right)^2 \ .
\end{equation}

Next, we consider the (generalized) driving force $F$ towards spontaneous curvature. Taking $H$ as constant over a small membrane patch,
\begin{equation}\label{eq:driving_force}
\begin{array}{rcl}
F & = & \frac{\partial w}{\partial H} \\[4mm]
& = & \frac{\partial}{\partial H} \left[ 2 \kappa \left( H - C_0 \right)^2 \right] \\[4mm]
& = & -4 \kappa \left( C_0 - H \right) \ .
\end{array}
\end{equation}
Eq.~(\ref{eq:driving_force}) resembles Hooke's law for a spring with spring constant $k = 4 \kappa$ and extension $u = \left( C_0 - H \right)$. The equilibrium position of the spring is $H = C_0$, which is a minimum as $\left( \partial^2 w / \partial H^2 \right) = 4 \kappa$ (which is positive for $\kappa > 0$). Note that it has been shown empirically that the average mean curvature $\left< H \right> = 0$ for bijels \cite{Lee2010AdvMater,Reeves2015InPrep}.

In order to understand how the driving force $F$ scales with particle size $r$, we first consider how the spontaneous curvature $C_0$ and the bending modulus $\kappa$ scale with $r$. $C_0$ has units of inverse length ($\mathrm{m}^{-1}$) and is expected to scale as $-1 / r$, which is backed up by analytical calculations for spherical particles on a spherical cap \cite{Kralchevsky2005Langmuir}. In that geometry, the result can also be explained using a scaling argument: to keep the angles fixed, including the particle's contact angle, both $r$ and the radius of curvature $R_{\mathrm{c}}$ of the spherical cap have to be reduced by the same factor, showing that
\begin{equation}\label{eq:spontaneous_curvature}
C_0 = -\frac{1}{R_{\mathrm{c}}} \propto -\frac{1}{r} \ .
\end{equation}
Note that $C_0$ also depends on the particle's contact angle $\theta$ and that $C_0 = 0$ for neutrally wetting particles ($\theta = 90^{\circ}$) \cite{Kralchevsky2005Langmuir}.

The bending modulus $\kappa$ has units of energy (J). As it is expected to depend on the W-L interfacial tension $\gamma_{\mathrm{WL}}$ (units $\mathrm{N}\:\mathrm{m}^{-1}$ ) and on the presence of the particles, one might guess
\begin{equation}\label{eq:bending_modulus}
\kappa \propto \gamma_{\mathrm{WL}} r^2 \ .
\end{equation}
This claim is backed up by analytical calculations of $\kappa$ for a close-packed monolayer of spherical particles on a spherical cap \cite{Kralchevsky2005Langmuir}.

In our experiments, the final bijel-channel width $L_{\mathrm{f}} \gg r$, so $\left| H \right| \sim 1 / L_{\mathrm{f}} \ll \left| C_0 \right|$ (Eq.~(\ref{eq:spontaneous_curvature})). Combined with Eqs.~(\ref{eq:driving_force}) and (\ref{eq:bending_modulus}), this means the driving force $F$ scales with $r$:
\begin{equation}\label{eq:driving_force_scaling}
\begin{array}{rcl}
F & = & -4 \kappa \left( C_0 - H \right) \\[4mm]
& \approx & -4 \kappa C_0 \\[4mm]
& \propto & -\gamma_{\mathrm{WL}} r^2 \cdot -\frac{1}{r} \\[4mm]
& \propto & \gamma_{\mathrm{WL}} r \ . \\[4mm]
\end{array}
\end{equation}
In words, for the same binary liquid ($\gamma_{\mathrm{WL}}$) and a given off-neutral wetting ($\theta \ne 90^{\circ}$), the driving force towards the spontaneous curvature is smaller for NPs than it is for MPs, which can help explain why fabricating bijels is possible over a larger range of heating rates with NPs than with MPs.

To gain a simple estimate of the disruption time $\Delta t_{\mathrm{d}}$, which is the time it takes for the driving force $F$ to cause so much curvature that bijel formation fails, we balance $F$ with a viscous drag force:
\begin{equation}\label{eq:balance_drive_viscosity}
\begin{array}{rcl}
F & = & F_{\mathrm{drag}} \\[4mm]
& \propto & \eta \lambda v \ , \\[4mm]
\end{array}
\end{equation}
where $\eta$ is viscosity, $\lambda \gg r$ the typical length scale of the disruption (independent of particle radius) and $v \sim \lambda / \Delta t_{\mathrm{d}}$. Combining Eqs.~(\ref{eq:driving_force_scaling}) and (\ref{eq:balance_drive_viscosity}), we get
\begin{equation}\label{eq:disruption_time_simple}
\Delta t_{\mathrm{d}} \propto \frac{\eta \lambda^2}{\gamma_{\mathrm{WL}} r} \ ,
\end{equation}
which is Eq.~(\ref{eq:disruption_time}).

Alternatively, consider the equation of motion of a damped oscillator (compare Eq.~(\ref{eq:driving_force})),
\begin{equation}\label{eq:overdamped_oscillator_eom}
m \ddot{u} + \mu \dot{u} + 4 \kappa u = 0 \ ,
\end{equation}
in which $\mu$ is a drag coefficient. We assume here that, at least initially, the drag mainly comes from the bulk fluids. In that case,
\begin{equation}\label{eq:drag_scaling}
\mu = \eta \lambda^3 \ .
\end{equation}
In our experiments $L_{\mathrm{f}} \gg r$, but if $L_{\mathrm{f}} \sim r$ then bulk drag may no longer dominate and effects of surface viscosity would have to be considered (which is outside of the scope of the current paper).

As the Reynolds number $\mathrm{Re} \ll 1$ here, even when considering motion at the scale of the channel width $L$, we can ignore the inertial term \cite{Russel1999}:
\begin{equation}\label{eq:overdamped_oscillator_eom_noninertial}
\mu \dot{u} + 4 \kappa u = 0 \ .
\end{equation}
Re-writing Eq.~(\ref{eq:overdamped_oscillator_eom_noninertial}) results in an expression for the rate of change of curvature $\left( \partial H / \partial t \right)$
\begin{equation}\label{eq:curvature_change_rate}
\begin{array}{rcl}
\dot{u} & = & -\frac{4 \kappa u}{\mu} \\[4mm]
\frac{\partial \left( C_0 - H \right)}{\partial t} & = & -\frac{4 \kappa \left( C_0 - H \right)}{\mu} \\[4mm]
\frac{\partial H}{\partial t} & = & \frac{4 \kappa \left( C_0 - H \right)}{\mu}
\end{array}
\end{equation}

Let us denote the time when the interfacial particles start interacting as $t_{\mathrm{in}}$. As at that time the bijel channel width $L \gg r$, we can write
\begin{equation}\label{eq:initial_curvature_rate}
\left( \frac{\partial H}{\partial t} \right)_{\mathrm{in}} \propto \frac{\kappa C_0}{\mu}  \ .
\end{equation}
For bijel disruption to occur, the curvature $H$ has to change by a threshold amount $\Delta H_{\mathrm{d}} \sim \lambda^{-1}$. For the disruption time, we can then write
\begin{equation}\label{eq:disrupt_time}
\begin{array}{lll}
\Delta t_{\mathrm{disrupt}} & \sim & \frac{\Delta H_{\mathrm{d}}}{\kappa C_0 / \mu} \\[4mm]
& \propto & \frac{\eta \lambda^2}{\gamma_{\mathrm{WL}} r} \ , \\[4mm]
\end{array}
\end{equation}
which is the same as Eq.~(\ref{eq:disruption_time_simple}). Interestingly, Eqs.~(\ref{eq:disruption_time_simple}) and (\ref{eq:disrupt_time}) suggest that lower quench rates could be used when using high-viscosity fluids (larger $\eta$). It has been reported that the binary liquid nitromethane-ethanediol is more forgiving in bijel fabrication than the W-L system (the viscosity of ethanediol is 16 times larger than for water) \cite{Tavacoli2011AdvFunctMater}.

\subsection{Jamming Time}\label{subsec:jamming_time}

Consider a bijel surface $S$ of area $A(t)$, i.e.~the area of the liquid-liquid interface between the two channels is decreasing during coarsening. Then the 2D packing fraction of particles on $S$ is
\begin{equation}{\label{eq:interface_packing_fraction}}
\phi(t) = \frac{N a_{\mathrm{WL}}(\theta)}{A(t)} \ ,
\end{equation}
with $a_{\mathrm{WL}}(\theta)$ the particle-interface cross-sectional area and $N$ the number of interfacial particles. Here, we assume that both $a_{\mathrm{WL}}(\theta)$ and $N$ are constant during the crucial (jamming) stages of bijel formation, for there is hardly any area left on $S$ for new particles to attach to. Eq.~(\ref{eq:interface_packing_fraction}) still holds for the bijel in its final i.e.~jammed state, so
\begin{equation}{\label{eq:interface_packing_final}}
\begin{array}{rcl}
\phi_{\mathrm{f}} & = & \frac{N a_{\mathrm{WL}}(\theta)}{A_{\mathrm{f}}} \\[4mm]
N a_{\mathrm{WL}}(\theta) & = & \phi_{\mathrm{f}} A_{\mathrm{f}} \ , \\[4mm]
\end{array}
\end{equation}
which leads to
\begin{equation}{\label{eq:interface_packing_in_final}}
\phi(t) = \phi_{\mathrm{f}} \frac{A_{\mathrm{f}}}{A(t)} \ . \\[4mm]
\end{equation}

As it is $L(t)$ rather than $A(t)$ that is typically reported from simulations and experiments, we write
\begin{equation}{\label{eq:bijel_VA}}
A(t) = c_{\mathrm{g}} \frac{V_{\mathrm{c}}}{L(t)} \ ,
\end{equation}
in which $c_{\mathrm{g}}$ is a geometrical pre-factor and $V_{\mathrm{c}}$ is the total volume of the bijel channel (which is constant during the phase separation of a symmetric binary liquid). Combining Eq.~(\ref{eq:interface_packing_final}) with (\ref{eq:bijel_VA}) gives
\begin{equation}{\label{eq:interface_packing_in_phi}}
\begin{array}{rcl}
\phi(t) & = & \phi_{\mathrm{f}} \frac{c_{\mathrm{g}} V_{\mathrm{c}}}{L_{\mathrm{f}}} \frac{L(t)}{c_{\mathrm{g}} V_{\mathrm{c}}} \\[4mm]
& = & \frac{\phi_{\mathrm{f}}}{L_{\mathrm{f}}} L(t) \ , \\[4mm]
\end{array}
\end{equation}
where we have assumed that $c_{\mathrm{g}}$ is constant i.e.~the topology of the bijel does not change substantially during the final stages of (successful) formation.

If $\phi_{\mathrm{in}}$ is the packing fraction at which interfacial particles start interacting, thereby affecting the phase separation \cite{Cheng2013JoCaIS}, then
\begin{equation}{\label{eq:jamming_time}}
\begin{array}{rcl}
\phi_{\mathrm{f}} - \phi_{\mathrm{in}} & = & \left( \frac{\phi_{\mathrm{f}}}{L_{\mathrm{f}}} \right) \int_{t_{\mathrm{in}}}^{t_{\mathrm{f}}} \frac{\mathrm{d}L}{\mathrm{d}t}\mathrm{d}t \\[4mm]
& \approx & \left( \frac{\phi_{\mathrm{f}}}{L_{\mathrm{f}}} \right) v_{L} \left( t_{\mathrm{f}} - t_{\mathrm{in}} \right) \\[4mm]
\Delta t_{\mathrm{j}} & = & t_{\mathrm{f}} - t_{\mathrm{in}} \approx \left( 1 - \frac{\phi_{\mathrm{in}}}{\phi_{\mathrm{f}}} \right) \left( \frac{L_{\mathrm{f}}}{v_{L}} \right) \ , \\[4mm]
\end{array}
\end{equation}
where in the second line we have used $L(t) \propto t$, which is valid in the relevant phase-separation regime for bijel formation (viscous-hydrodynamic) \cite{Pagonabarraga2001NewJournalPhysics}.

Note that Eq.~(\ref{eq:jamming_time}) can explain several observations. First, the larger $L_{\mathrm{f}}$, the longer the jamming time, which may help explain the empirical upper limit to bijel channel width \cite{Witt2013SoftMatter}. Secondly, the larger the coarsening speed $v_{L}$, the shorter the jamming time. As $v_{L}$ increases with heating rate, through its dependence on the temperature-dependent interfacial tension \cite{Bray2002AiP}, this may help explain why heating faster facilitates successful bijel formation (even for MPs).

\providecommand{\noopsort}[1]{}\providecommand{\singleletter}[1]{#1}%

\end{document}